# Electrostatic Structures in Space Plasmas: Stability of Two-dimensional Magnetic Bernstein-Greene-Kruskal Modes

C. S. Ng, S. J. Soundararajan, and E. Yasin

*Geophysical Institute, University of Alaska Fairbanks, Fairbanks, Alaska 99775, USA*

**Abstract.** Electrostatic structures have been observed in many regions of space plasmas, including the solar wind, the magnetosphere, the auroral acceleration region, and in association with shocks, turbulence, and magnetic reconnection. Due to potentially large amplitude of electric fields within these structures, their effects on particle heating, scattering, or acceleration can be important. One possible theoretical description of some of these structures is the concept of Bernstein-Greene-Kruskal (BGK) modes, which are exact nonlinear solutions of the Vlasov-Poisson system of equations in collisionless kinetic theory. BGK modes have been studied extensively for many decades, predominately in one dimension (1D), although there have been observations showing that some of these structures have clear 3D features. While there have been approximate solutions of higher dimensional BGK modes, an exact 3D BGK mode solution in a finite magnetic field has not been found yet. Recently we have constructed exact solutions of 2D BGK modes in a magnetized plasma with finite magnetic field strength in order to gain insights of the ultimate 3D theory [Ng, Bhattacharjee, and Skiff, Phys. Plasmas **13**, 055903 (2006)]. Based on the analytic form of these solutions, as well as Particle-in-Cell (PIC) simulations, we will present numerical studies of their stability for different levels of background magnetic field strength.

**Keywords:** Electrostatic structures, BGK modes, Solitons, Plasma kinetic equations.

**PACS:** 52.35.Sb, 05.45.Yv, 52.25.Dg, 94.30.Kq.

## INTRODUCTION

A BGK mode is an exact steady-state nonlinear solution of the Vlasov-Poisson system first found by Bernstein, Greene, and Kruskal [1]. Since then, many papers have been written on this subject. However, the vast majority of these works were still within the original 1D framework. BGK modes have been observed experimentally in both magnetized plasmas, and trapped pure electron plasmas. There is also a long history of space observations of electrostatic structures (with descriptions like solitary waves or phase space holes) in the solar wind, the magnetosphere, the auroral acceleration region, and in association with shocks, turbulence, and magnetic reconnection, with some of them considered as BGK modes, e.g., [2,3].

More recently there has been renewed interest in 3D BGK modes. This is mainly due to 3D features of solitary wave structures in space-based observations that cannot be explained by 1D BGK modes. For example, electrostatic solitary waves in the auroral ionosphere were observed as having electric field components perpendicular to the background magnetic field with comparable magnitude to the parallel component

[4]. This is inconsistent with a 1D BGK-like potential, which only has parallel component, but is consistent with the structure of a single-humped solitary potential that travels past the spacecraft along the magnetic field. It was not until recently that 3D BGK modes have been constructed under the assumption of an infinitely strong background magnetic field [5]. The strong field assumption constrains charged particles to move along magnetic field lines, and thus effectively reduces the problem to 1D. Efforts on relaxing this assumption has obtained limited success so far. In our recent work, we have found exact symmetric 3D solution in the unmagnetized (**B** = 0) case [6], as well as exact 2D solution for finite magnetic field [7], by using distribution function $f$ that depends not only on energy, but also on angular momentum.

While we have constructed these higher dimensional BGK modes, whether such solutions can be realized and are stable enough to be observable remains an open question. In this paper, we will present our first effort in studying the stability of the 2D BGK mode based on Particle-in-Cell (PIC) simulations. Note that this problem is different from previous stability study of 1D BGK mode in higher dimensions, e.g., [8,9,10], since our study is on a BGK mode that is localized in 2D with cylindrical symmetry, which is very different as compared with the 1D BGK mode.

In the following, we describe the functional form of our 2D BGK mode solution and how we setup the PIC simulations. Then we describe numerical results obtained so far and discuss the stability of cases with different magnetic field levels.

## NUMERICAL SETUP AND RESULTS

Following the derivation in Ref. [7], the 2D BGK mode considered here is characterized by an electrostatic potential $\psi = \psi(\rho)$ that depends only on the radial coordinate of a cylindrical coordinate system with a static uniform magnetic field $\mathbf{B} = B_0 \hat{\mathbf{z}}$ along the $z$-axis. To satisfy the steady state Vlasov equation, the electron distribution function is of the form $f = f(w,l)$, with $w = v^2/2 - \psi = \left(v_\rho^2 + v_\phi^2 + v_z^2\right)/2 - \psi$ being the energy, and $l = 2\rho v_\phi - B_0 \rho^2$ being two times the generalized angular momentum. Note that we have expressed quantities in dimensionless form with velocity $v$ measured in electron thermal velocity $v_e$, length scale in electron Deybe length $\lambda = v_e/\omega_{pe} = (v_e/e)\sqrt{\varepsilon_0 m_e/n_0}$ (with $e$, $m_e$, $n_0$ electron charge, mass and density respectively), $\psi$ in $n_0 e \lambda^2/\varepsilon_0$, $f$ in $n_0/v_e^3$, energy in $m_e v_e^2$, magnetic field in $n_0 e \lambda/\varepsilon_0 v_e$ so that the normalized magnetic field is simply the ratio between the electron cyclotron frequency $\omega_{ce} = eB_0/m_e$ and the plasma frequency $\omega_{pe}$. The ion distribution is assumed to provide a uniform background of positive charge, which is simply of the value of unity in the above normalization scheme. Note that this assumption is mainly for simplicity and can be relaxed with small changes in the final results as long as we are considering the case with a positive $\psi$. With cylindrical symmetry in these forms, it can be shown by direct substitution that $f$ satisfies the steady state Vlasov equation of the form

$$v_\rho \frac{\partial f}{\partial \rho} + \left(\frac{d\psi}{d\rho} + \frac{v_\phi^2}{\rho} - B_0 v_\phi\right)\frac{\partial f}{\partial v_\rho} - \left(\frac{v_\rho v_\phi}{\rho} - B_0 v_\rho\right)\frac{\partial f}{\partial v_\phi} = 0. \qquad (1)$$

A particular choice of the form of $f$ is

$$f(w,l) = (2\pi)^{-3/2} \exp(-w)\left[1 - h_0 \exp(-kl^2)\right], \qquad (2)$$

where the two constant parameters satisfy $1 > h_0 > 0$, and $k > 0$. In order that the electric potential is obtained from this form of $f$, $\psi$ has to satisfy the Poisson equation,

$$\frac{1}{\rho}\frac{d}{d\rho}\left(\rho\frac{d\psi}{d\rho}\right) = e^{\psi(\rho)}\left[1 - \frac{h_0}{\sqrt{1+8k\rho^2}}\exp\left(-\frac{kB_0^2\rho^4}{1+8k\rho^2}\right)\right] - 1, \qquad (3)$$

after integrating $f$ of the form of Eq. (2). Eq. (3) can be solved numerically by a scheme described in Ref. [7] with the boundary condition that $\psi(\rho \to \infty) \to 0$ for a localized solution. Note also that the above solution can be generalized to satisfy the Ampère's law also, with the solution very close to the one described above if $v_e$ is much smaller than the speed of light as assumed here.

The 2D PIC code we used for this study is a commercially available code OOPIC Pro (txcorp.com). In this first study, the code is run on a serial computer and thus it is more limited in resolutions. To study the stability of a BGK mode solution, electrons are assigned initially according to Eq. (2) with $\psi$ given by solving Eq. (3). Ions are treated as immobile in this study in consistent with the assumption used in deriving the exact solution. Since periodic boundary condition is much easier to handle in the PIC code, the solution is put at the center of a 2D doubly periodic square in the $x$-$y$ plane that is much larger than the size of the electric potential. This is possible since for large distance compared with the size of the 2D BGK mode, the distribution tends to simply a Maxwellian with uniform density. In fact, by convenience of setting the initial condition, the lowest moments (density $n_e$, average velocity $\langle \mathbf{v} \rangle$, and temperature $T_e$) of the electron distribution $f$ given by Eq. (2) is assigned for each cell. These quantities can be calculated by expressions derived from direct integrations,

$$n_e = e^{\psi(\rho)}\left[1 - \frac{h_0}{\sqrt{1+8k\rho^2}}\exp\left(-\frac{kB_0^2\rho^4}{1+8k\rho^2}\right)\right] \qquad (4)$$

$$\langle \mathbf{v} \rangle = -\frac{\hat{\phi}}{n_e}\frac{4h_0 kB_0\rho^3}{(1+8k\rho^2)^{3/2}}\exp\left(\psi - \frac{kB_0^2\rho^4}{1+8k\rho^2}\right) \qquad (5)$$

$$T_e = \frac{1}{3}\left(2 + \frac{e^\psi}{n_e}\left\{1 - h_0\left[\frac{1+8k\rho^2+(4kB_0\rho^3)^2}{(1+8k\rho^2)^{5/2}}\right]\exp\left[-\frac{kB_0^2\rho^4}{1+8k\rho^2}\right]\right\} - \langle \mathbf{v} \rangle^2\right) \qquad (6)$$

where $T_e$ is in the unit of $m_e v_e^2/k_B$. Specifying these moments for each cell give a good representation of the distribution for the study of the stability of the BGK mode. For example, Fig. 1 shows profiles of $\psi$, $n_e$, $\langle v_\phi \rangle$, and $T_e$ as functions of $\rho$ for a case with $h_0 = 0.9$, $k = 0.1$, and $B_0 = 1$. We have performed simulations using the same $h_0$ and $k$ for different levels of $B_0 = 10, 1$, and $0.2$. Moreover, to provide control cases for comparisons to show if BGK mode solutions are indeed special cases, simulations are also done using initial conditions the same as the BGK mode cases except with reversing $\langle v_\phi \rangle$ as given by Eq. (5). We will call these non-BGK mode cases in the following. These simulation results are easier to comprehend by viewing movies made from these runs. In this paper, we can only provide descriptions of these runs with the help of some still figures. Interested readers can view these movies, with the help of a presentation file, available in "http://www.gi.alaska.edu/~chungsangng/bgk/".

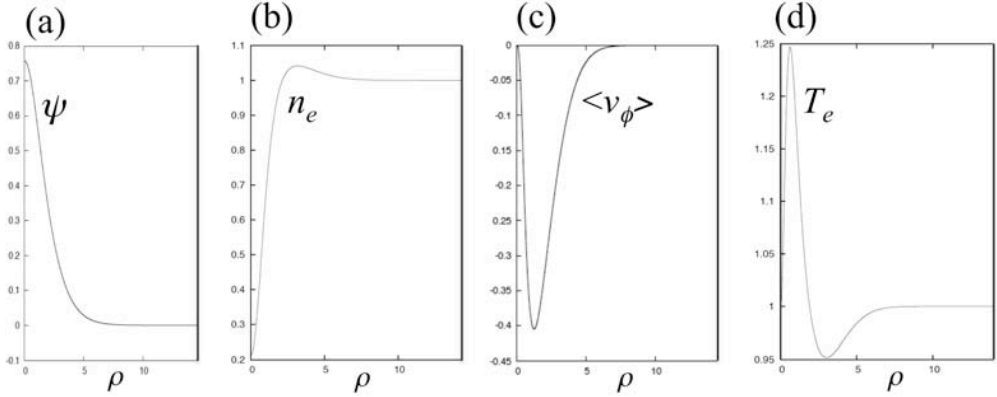

**FIGURE 1.** $\psi$, $n_e$, $\langle v_\phi \rangle$, and $T_e$ as functions of $\rho$ for a case with $h_0 = 0.9$, $k = 0.1$, and $B_0 = 1$.

For the case with $B_0 = 10$, the size of the solution is smaller than that of the $B_0 = 1$ case and thus a smaller simulation box is used. The electron density close to the initial moment ($t = 0.1\tau_{ce}$, where $\tau_{ce} = 2\pi/\omega_{ce}$ is the electron cyclotron period) is shown in Fig. 2(a) in the filled color coded contours with the red part at the center represents $n_e \sim 0.2$ and the green part outside represents $n_e \sim 1$, the asymptotic average level, and thus blue represents $n_e$ enhanced to greater than unity. Note that only the central region of the simulation box is plotted so as to show clearer the central structure. The simulation box for this case is $-15 < x, y < 15$. There is $160^2$ number of cells with a total of $7.44 \times 10^5$ numerical particles. During the whole period of this run up to 50 $\tau_{ce}$, the $n_e$ profile actually does not change very much from Fig. 2(a). By more careful inspection of electric field profile (not shown), small oscillations (up to ~ 10% of the peak electric field) are observed with a period about one $\tau_{ce}$.

For the non-BGK mode run with $B_0 = 10$, the initial $n_e$ profile is very similar to Fig. 2(a) since only the average velocity field is reversed initially. However, much stronger oscillations are observed to appear right away with the same $\tau_{ce}$ period for the same simulation duration as the BGK mode run. The $n_e$ profile alternates between Fig. 2(a) (with the peak electric field about 20% less than that of the BGK mode) and 2(b) (taken at $t = 0.6\tau_{ce}$) when the central $n_e$ has a much smaller deviation from unity (with even some enhanced of $n_e$ at same isolated spots, the blue spots, a little away from the center), and the peak electric field is much smaller (~ half of the initial value). From these two runs, we have confirmed for the first time that the 2D BGK mode found in Ref. [7] is indeed stable in the strong magnetic field limit. This is because an initial condition close to the exact BGK mode solution (numerically it is not possible to impose exact initial condition) would remain close to steady in time with small fluctuations of frequency close to $\omega_{ce}$, while this is not true for the run with non-BGK initial condition. However, it is also interesting to see that the non-BGK mode structure in the strong field limit does not simply disintegrate or decay away in time. Instead, it keeps the overall electric field structure only with large oscillations, again in the frequency close to $\omega_{ce}$.

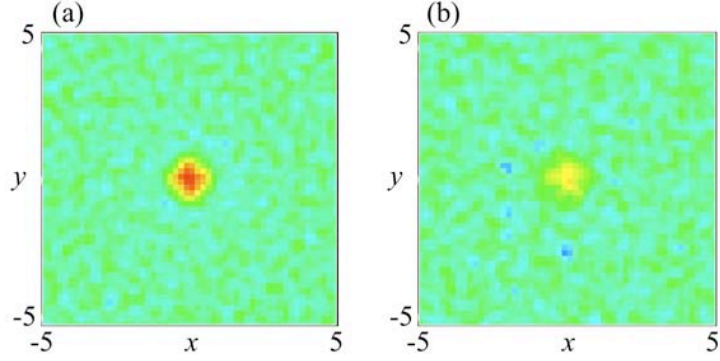

**FIGURE 2.** (a) Color coded 2D profile of $n_e$ (see text for descriptions) for a run with $B_0 = 10$ for a BGK mode initial condition at $t = 0.1\,\tau_{ce}$. (b) $n_e$ for the non-BGK run with $B_0 = 10$ at $t = 0.6\,\tau_{ce}$.

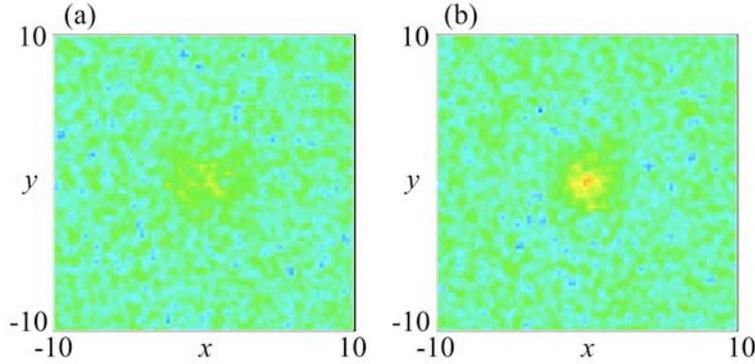

**FIGURE 3.** (a) Color coded 2D profile of $n_e$ for a run with $B_0 = 1$ for a BGK mode initial condition at $t = 0.305\,\tau_{ce}$. (b) $n_e$ for the same run at $t = 6.65\,\tau_{ce}$.

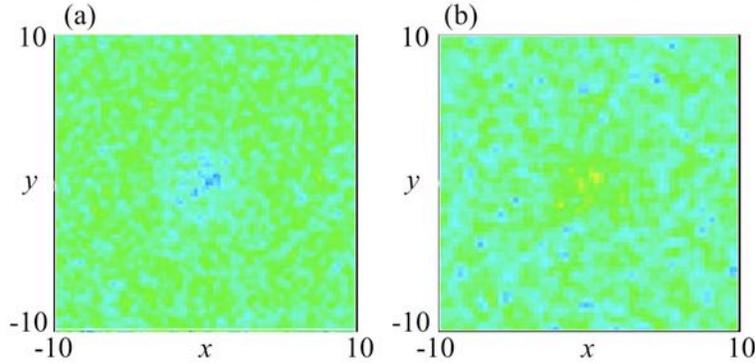

**FIGURE 4.** (a) Color coded 2D profile of $n_e$ for a run with $B_0 = 1$ for a non-BGK mode initial condition at $t = 1.2\,\tau_{ce}$. (b) $n_e$ for a run with $B_0 = 0.2$ for a BGK mode initial condition at $t = 1.3\,\tau_{ce}$.

The case with a moderate magnetic field strength is very difference from the strong field case. For a run with BGK mode initial condition with $B_0 = 1$, while the initial $n_e$ profile is similar to Fig. 2(a), the density structure quickly dissolve in a fraction of $\tau_{ce}$, with an associated decay of the electric field structure, as shown in Fig. 3(a) taken at $t = 0.305\,\tau_{ce}$. As shown in the plot, the density depletion is reduced and also fragmented, although such a decreased density structure still produced an overall outward electric field with strength much reduced from the initial condition (~50%). However, the structure does come back at roughly one $\tau_{ce}$. This process repeats itself quasi-periodically for many $\tau_{ce}$. The $n_e$ profile at a point with roughly maximum depletion as shown in Fig. 3(b) taken at $t = 6.65\tau_{ce}$ when the density structure reappears the eighth times. Therefore the oscillation period is a little shorter than $\tau_{ce}$ (about 83%). We see that the BGK mode for the $B_0 = 1$ case is much less stable than the $B_0 = 10$ case in the sense that the fluctuations is much stronger, although a structure similar to the initial profile does come back periodically, similar to the non-BGK mode for the $B_0 = 10$, except with even more finer and rapid fluctuations within a period. Nevertheless, the BGK mode for the $B_0 = 1$ case still keeps its structure better than the non-BGK mode for the same magnetic field. First, the initial disintegration for the non-BGK mode is more complete in the sense that the initial electric field structure is soon dissolved into noise level. The structure does reappear in a much quicker manner at around one $\tau_{ce}$ but with a stronger oscillation that the central $n_e$ quickly get back to unity, after coming back to the maximum depletion level, and then overshoot briefly to a level greater than unity. Fig. 4(a) shows $n_e$ profile for the non-BGK mode run at $t = 1.2\tau_{ce}$ when this happens. When the magnetic field level is decreased further, the BGK mode becomes even more unstable. A run with $B_0 = 0.2$ shows that the original structure is disintegrated and not recovered even after a few $\tau_{ce}$, although $n_e$ still has an overall slight depletion in the central region with random fluctuations of density fragments of time scales much shorter than $\tau_{ce}$. Fig. 4(b) shows an instant at $t = 1.2\tau_{ce}$ when such fluctuations are stronger.

## CONCLUSION

We have presented results of our first effort in studying the stability of 2D BGK modes based on PIC simulations. It is confirmed that an initial condition using the analytic solution does stay roughly steady in time and thus is stable in the strong magnetic field limit. However, non-BGK initial condition also keeps overall structure except with strong oscillations in a time scale of about $\tau_{ce}$. BGK modes are found to be progressively less stable with weaker magnetic field strength.

## ACKNOWLEDGMENTS

This work is supported by a National Science Foundation grant PHY-1004357.